%% file: DimVis.tex
\documentclass{egpubl}
\usepackage{mlvis2024}

 \WsPaper           

\usepackage[T1]{fontenc}
\usepackage{dfadobe}
\usepackage{tabularx}
\usepackage{multirow}
\usepackage{graphicx} 
\usepackage{amssymb} 
\usepackage{makecell}
\usepackage{float}
\usepackage{t1enc}

\usepackage{cite}  
\electronicVersion
\PrintedOrElectronic
\ifpdf
  \pdfoutput=1\relax                   
  \usepackage{graphicx}                
  \DeclareGraphicsExtensions{.pdf,.png,.jpg,.jpeg} 
\else
  \usepackage{graphicx}                
  \DeclareGraphicsExtensions{.eps}     
\fi%

\usepackage{egweblnk}

\usepackage[dvipsnames]{xcolor} 

\graphicspath{{figures/}{pictures/}{images/}{./}} 

\usepackage{cleveref}
\renewcommand{\autoref}{\Cref}

\usepackage{atbegshi}
\usepackage[absolute]{textpos}
\AtBeginShipout{%
   \begin{textblock*}{3mm}(105mm,280mm)
	\textnormal{\the\numexpr\value{page}}
   \end{textblock*}%
}

\title[DimVis: Interpreting Visual Clusters in Dimensionality Reduction With Explainable Boosting Machine]%
      {DimVis: Interpreting Visual Clusters in Dimensionality Reduction With Explainable Boosting Machine}

\author[Salmanian et al.]
{\parbox{\textwidth}{\centering \vspace{-1.7cm} P. Salmanian$^{1,3}$\orcid{0000-0001-7719-127X}, A. Chatzimparmpas$^{2}$\orcid{0000-0002-9079-2376}, A. C. Karaca$^{3}$\orcid{0000-0002-6835-7634}, and R.\,M. Martins$^{1}$\orcid{0000-0002-2901-935X}}
        \\
{\parbox{\textwidth}{\centering \vspace{-1.8cm} $^1$Department of Computer Science and Media Technology, Linnaeus University, Sweden\\
$^2$Department of Information and Computing Sciences, Utrecht University, The Netherlands\\
$^3$Department of Computer Engineering, Yildiz Technical University, Turkey\\
}}
}

\teaser{
 \centering \vspace{-14mm}
 \includegraphics[width=0.88\linewidth]{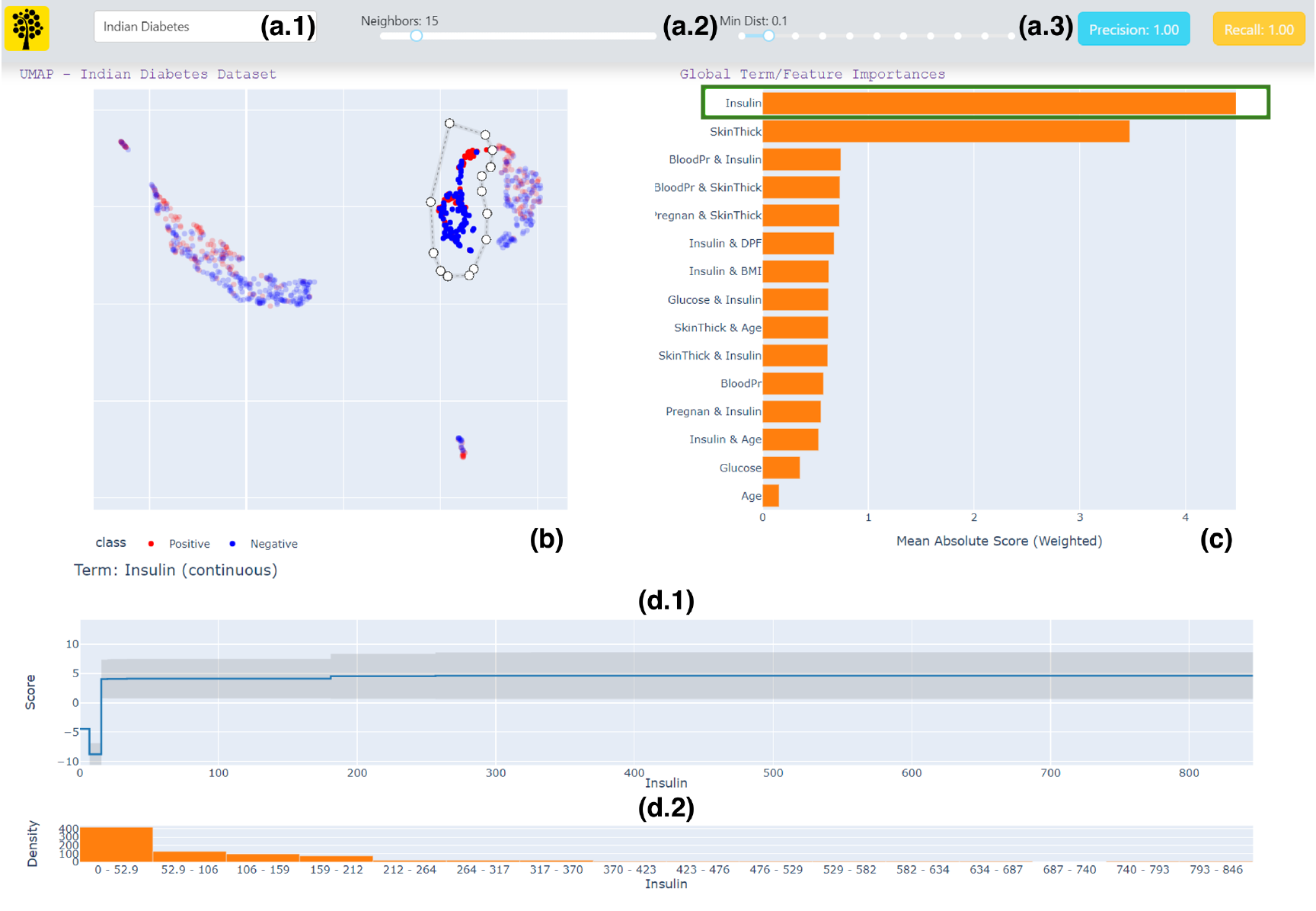} \vspace{-5mm}
 \caption{Components of \textsc{DimVis}: a panel for (a.1) choosing a dataset and (a.2) adjusting UMAP hyperparameters, and two performance metrics (a.3) 
 related to the underlying EBM model that supports the visual exploration;
 (b) a UMAP projection of the dataset, with a selection of points that determine a visual cluster of interest; 
 (c) shows a ranking of features (single and pairs) for the selected visual cluster, based on the EBM model's feature importance; 
 and (d.1) presents the Score of the selection, while (d.2) shows the distribution for one feature.
 }
 \label{fig:teaser}
}

\begin{document}

\maketitle

\begin{abstract}
   \input{0.Abstract}

\end{abstract}

\section{Introduction} \label{sec:intro}
	\input{1.Introduction}

\section{Related Work} \label{sec:relwo}
	\input{2.Related_work}

\section{\textsc{DimVis}: System Overview and Use Case} \label{sec:overview}

\input{3.System_overview}

\section{Usage Scenario: Interpreting Clusters in a Projection} \label{sec:case}
	\input{4.Use_case}

\section{Discussion} \label{sec:disc}
	\input{5.Discussion}

\section{Conclusion} \label{sec:con}
	\input{6.Conclusion}

\bibliographystyle{eg-alpha-doi}

\bibliography{DimVis.bib}

\end{document}

%% file: 0.Abstract.tex
Dimensionality Reduction (DR) techniques such as t-SNE and UMAP are popular for transforming complex datasets into simpler visual representations. However, while effective in uncovering general dataset patterns, these methods may introduce artifacts and suffer from interpretability issues. This paper presents DimVis, a visualization tool that employs supervised Explainable Boosting Machine (EBM) models (trained on user-selected data of interest) as an interpretation assistant for DR projections. Our tool facilitates high-dimensional data analysis by providing an interpretation of feature relevance in visual clusters through interactive exploration of UMAP projections. Specifically, DimVis uses a contrastive EBM model that is trained in real time to differentiate between the data inside and outside a cluster of interest. Taking advantage of the inherent explainable nature of the EBM, we then use this model to interpret the cluster itself via single and pairwise feature comparisons in a ranking based on the EBM model's feature importance. The applicability and effectiveness of DimVis are demonstrated via a use case and a usage scenario with real-world data. We also discuss the limitations and potential directions for future research.

   \makeatletter

	\def\customclassification{\vskip 5.5pt\par\reset@font\rmfamily}
	\def\endcustomclassification{\relax}
	\makeatother

	\begin{customclassification}
		\textbf{CCS Concepts}\\
		$\bullet$ \textbf{Human-centered computing} $\rightarrow$ Visualization; Visual analytics; 
		$\bullet$ \textbf{Machine learning} $\rightarrow$ Unsupervised learning;
	\end{customclassification}

%% file: 1.Introduction.tex
Dimensionality Reduction (DR) is an important component of the visual exploration of complex and high-dimensional datasets, transforming them into more manageable (but still representative) lower-dimensional forms with the aim to facilitate interpretation and improve computational efficiency while retaining essential information for analysis~\cite{Sacha2017Visual}. 
However, the outcomes of modern nonlinear DR techniques such as t-SNE~\cite{maaten2008visualizing} and UMAP~\cite{mcinnes2018umap} can be challenging to interpret due to the lack of inherent meaning to the extracted shapes/clusters, suboptimal hyperparameters (e.g., \cite{Wattenberg2016How}), and misleading distortions introduced during the process~\cite{heulot2017visualizing,nonato2018multidimensional}.
To address these challenges, some visualization tools have been developed with the goal of inspecting DR projections interactively to assess the global and local quality of the 2D space~\cite{Chatzimparmpas2020A,Chatzimparmpas2020The,Chatzimparmpas2024Visualization} and to produce meaningful interpretations to the extracted shapes and clusters~\cite{fujiwara2019supporting,marcilio2021}. One recent example from Bibal et al.~\cite{bibal2021ixvc,bibal2023dtsne} involved training an interpretable supervised Machine Learning (ML) model, the decision tree (DT), and use it to interpret DR layouts. However, DTs suffer from scalability issues, getting quickly deep and unparsable, and may overfit the data to achieve high fidelity. One way to overcome these limitations is by using the Explainable Boosting Machine (EBM)~\cite{Lou2013Accurate}, a tree-based, cyclic gradient boosting generalized additive model, which is a powerful alternative that remains interpretable and scalable even with large-scale, high-dimensional datasets~\cite{caruana2015intelligible}.

We propose a new visualization tool, \textsc{DimVis}, which employs EBM models for unveiling the factors influencing cluster formation and data relationships in nonlinear projections. It enables users to 
interactively inspect visual patterns of the DR layout (clusters, shapes, etc.) to gain insights about their content by training an EBM model on-the-fly on user-selected data points. The explanation is based on exploring single and pairs of features ranked by EBM's intrinsic feature importance regarding their ability to explain the separation between the selection and the remaining points of the dataset. The results are visualized with intuitive and simple visualizations such as bar charts, line plots, histograms, and heatmaps. We demonstrate \textsc{DimVis} via a use case and a usage scenario with healthcare datasets (also used, for example, in \cite{Ming2019RuleMatrix} and \cite{chatzimparmpas2020tvisne}), showcasing the tool's capability of analyzing high-dimensional datasets and enhancing the interpretability of the resulting 2D space from a DR technique.

%% file: 2.Related_work.tex

In early research on DR interpretation, static visualizations were generated by considering the layout as a whole, without giving the users the ability and flexibility to specify their own regions of interest~\cite{dasilva2015attribute}. Interactive interpretation of selections and groups, as is the case for \textsc{DimVis}, has been explored for example in Probing Projections~\cite{stahnke2016}, where groups of points (obtained either by selection or a clustering algorithm) are compared to each other or to the rest of the dataset via individual histograms for each dimension; the users then must visually determine dimension importance by themselves. This approach provides a strong foundation but may present problems with very high-dimensional datasets, and the interpretations are inherently univariate.

Some examples of recent work that are more related to \textsc{DimVis} are those from Marcilio-Jr et al.~\cite{marcilio2021} and Fujiwara et al.~\cite{fujiwara2019supporting}. In both cases, users can define groups of points of interest and interpret them by contrasting their features against outside points. Fujiwara et al. use an adapted version of contrastive PCA~\cite{cPCA}, which in itself is an adapted version of the classic PCA for highlighting the differences between two sets of data points. Although relevant, we argue that being a linear algorithm, interpretations based on adapted versions of PCA will still have limitations when dealing with very high-dimensional and complex groups of points. Marcilio-Jr et al.~\cite{marcilio2021} used Shapley values to generate explanations for each data point based on its distance to the centroids of the specified clusters, which are then visualized with a relatively complex visual abstraction based on merging histograms and connecting them with edges (as lines). In contrast, \textsc{DimVis} uses basic, simple visualizations that are arguably more natural and easily interpreted, and the explanations come directly from an inherently interpretable and robust supervised model instead of being generated with a black-box, model-agnostic method.

DT-SNE~\cite{bibal2023dtsne} is a simpler but \emph{static} approach combining t-SNE with supervised decision trees for interpreting projections with features, and IXVC~\cite{bibal2021ixvc} supports users in interactively explaining projections with decision trees. However, decision trees, especially visualized as node-link diagrams, can become quickly hard to interpret, and the increase in the number of dataset features may heavily affect their scalability. On the contrary, \textsc{DimVis} uses EBM -- which can be as powerful as ensemble learning models -- and offers simple and scalable visual representations for exploring the most important features in user-selected data contrastively.



%% file: 3.System_overview.tex
\textsc{DimVis} is an open-source, web-based visualization tool~\cite{Parisa2023DimVis} 
which uses the state-of-the-art, supervised ``glass-box'' EBM model~\cite{Lou2013Accurate} to interpret visualizations generated with unsupervised DR techniques. We chose the EBM because its results and pre-defined visualizations (which we used) have already been successfully evaluated with real-world domain experts~\cite{caruana2015intelligible}.

After selecting the dataset (shown in \autoref{fig:teaser}(a.1)), 
initially only views indicated in (a.1), (a.2), and (b) are shown. To initiate the exploration and trigger the remaining views, a lasso selection of data points must be made in the UMAP view (b) containing a visual cluster perceived by the user as being of interest. 
%
As soon as a selection is made, an EBM model is trained in the background (more details below) and the remaining visualizations are presented. The model's precision and recall are visible (in~\autoref{fig:teaser}(a.3)), as well as single and pairwise feature importances extracted from the EBM model and visualized in a bar chart sorted in descending order from top to bottom (see~\autoref{fig:teaser}(c)). On click, users can explore further a feature with a Score line plot (\autoref{fig:teaser}(d.1)) and a Density histogram (\autoref{fig:teaser}(d.2)), or two features with a Score heatmap.



\begin{figure*}[ht]
 \centering 
 \includegraphics[width=\linewidth] {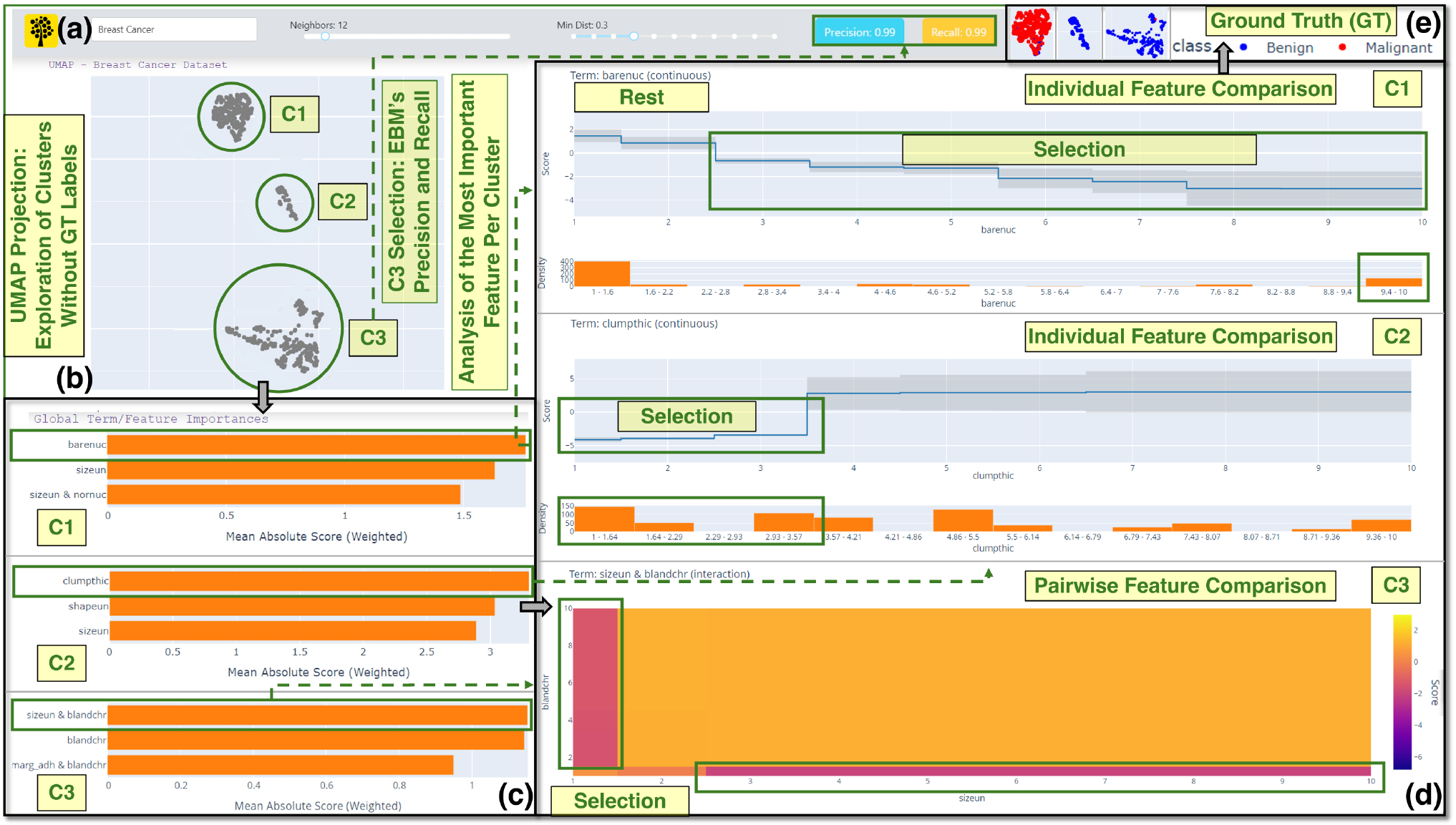}
 \vspace{-7mm}
 \caption{Exploration of three clusters (C1--C3) produced by UMAP in (b) and local Feature Importances for the top three important features of the EBM model (c), trained on each cluster. (a) shows the dataset, UMAP hyperparameters, and EBM's precision and recall for C3. The most important feature (pair) per cluster is further explained in (d), with the selection mapped with a negative Score and the remaining points a positive one. For a single feature, Density histograms show the distribution of values. (e) reveals the GT labels used solely for verification.}
 \label{fig:interactive} \vspace{-2mm}
 \end{figure*}

We further explain the tool by introducing a first use case with the \emph{Breast Cancer Wisconsin (Original)} dataset (entries: 699; dimensions: 9 (graded 1--10); classes: 2) obtained from the UCI ML repository~\cite{Dua2017Machine}. The class labels (\emph{Benign} or \emph{Malignant}) are only used as a final verification for the insights (Figure~\ref{fig:interactive}(e)).

\textsc{DimVis} utilizes the UMAP algorithm~\cite{mcinnes2018umap} 
and users can interactively adjust UMAP's hyperparameters -- ``Number of Neighbors'' (Neighbors) and ``Minimum Distance'' (Min Dist) -- to explore different projections (see \autoref{fig:interactive}(a)). The projection in~\autoref{fig:interactive}(b) contains three distinct clusters (C1--C3), and no GT labels to simulate a typical unsupervised learning scenario.
%
When a group of points is selected with the lasso functionality, 
the EBM gets retrained and selected points are marked with class $0$ while the remaining points are marked with class $1$. 
Using the inherent interpretability features of the EBM model, we rank the main features (single and pairs) that contribute to the separation of the cluster according to the trained model, using the Mean Absolute Score (Weighted). As shown in \autoref{fig:interactive}(c), the most important single and pair of features are: the number of bare nuclei for C1 (\textsc{barenuc}, $\approx$$1.7$), clump thickness for C2 (\textsc{clumpthic}, $\approx$$3.2$), and the combination of uniformity of cell size and bland chromatin for C3 (\textsc{sizeun \& blandchr}, $\approx$$1.2$). That can be further explored in relation to the actual values and the distribution of the data entries (see below). More algorithmic details regarding EBM can also be found in Lou et al.~\cite{Lou2013Accurate}.

When a user clicks on a single feature in the bar chart (\autoref{fig:interactive}(c), C1), a line plot and a histogram appear, displaying the impact of that specific feature (\textsc{barenuc}, in this case) on the model's predictions. As illustrated in \autoref{fig:interactive}(d), C1 (top), the line plot shows the \emph{Score}, i.e., the contribution of each feature's value inside (\emph{negative} values) or outside (\emph{positive} values) of the selection. It also shows the \emph{quantified uncertainty}, that is, the variance in the final predictions, encoded as gray bands. For C2, for example, we can see that values of \textsc{clumpthic} up to 3.5 are more related to the points inside the selected cluster, with high certainty. The rest of the values are more related to the points outside the cluster, although with less certainty.
The histogram depicts the \emph{Density}, which is a simple distribution of the values for a specific feature across all points globally. 
For C1, the feature \textsc{barenuc} has a clear separation between low and high values that seems to be related to the cluster formation, while for C2, \textsc{clumpthic} does not seem to show any clear pattern for points inside or outside the cluster (which shows why it is important to consider this view together with the \emph{Score} line plot).
%
%
%
%
%
Additionally, \textsc{DimVis} allows users to explore pairs of features, as shown in~\autoref{fig:interactive}(d). The heatmap uses color intensity to represent the level of interaction (i.e., from purple for \emph{negative Score} to yellow for \emph{positive Score}). 
In~\autoref{fig:interactive}(d), C3, we see that points where one of the two features is low while the other is high are more related to the cluster (see green boxes); when both features are low, or both are high, then the \emph{Score} is higher, indicating a stronger relation to the points outside the cluster.


In the next section we explore different aspects of the user experience of \textsc{DimVis} by simulating a usage scenario in the domain of healthcare, one of the important application areas of DR.

\begin{figure*}[t]
 \centering
 \includegraphics[width=\linewidth]{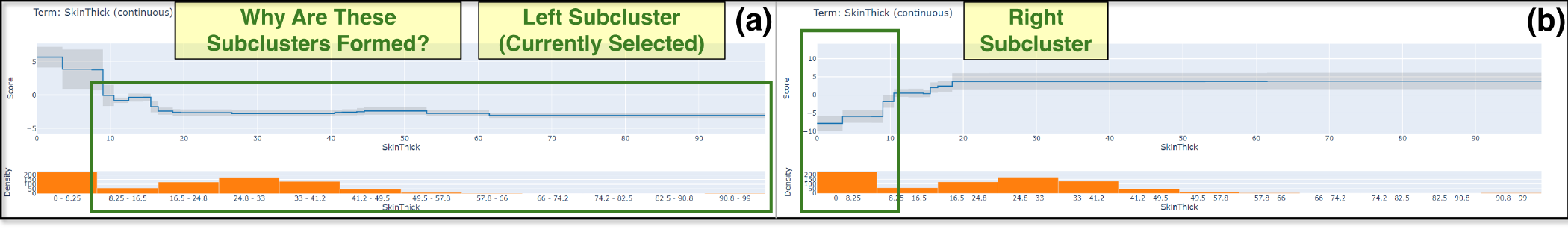}
 \vspace{-7mm}
 \caption{The individual feature importance of SkinThick explains why the user-selected left and right subclusters in (a) and (b), respectively, were formed. The left subcluster (negative Score) contains points mostly with moderate values for SkinThick (Density) and low uncertainty (see the tight, gray band), while the right subcluster includes points with low values for SkinThick and relatively moderate uncertainty.}
 \label{fig:subclusters} \vspace{-2mm}
 \end{figure*}

%% file: 4.Use_case.tex
For this hypothetical scenario, we follow a physician specializing in diabetes (which we will call Sara). She recently learned about UMAP but, while enthusiastic, she remains doubtful about the reliability of such projections. Hence, she turns to \textsc{DimVis} for an in-depth analysis of UMAP projections. Sara uploads to \textsc{DimVis} (\autoref{fig:teaser}(a.1)) the \emph{Pima Indian diabetes} dataset (entries: 768; dimensions: 8; classes: 2) from the UCI ML repository~\cite{Dua2017Machine}.


Without any selection, \textsc{DimVis} trains a supervised EBM model (precision: $0.81$; recall: $0.82$) and visualizes the Global Feature Importances from the most to the least important: Glucose, BMI, Age, DPF, Insulin, Pregnan, SkinThick, and BloodPr (not shown due to space limits).
%
Now that she has gained an overview, Sara wants to investigate why a cluster in \autoref{fig:teaser}(b) is split into left and right subclusters. By using a lasso to choose the left subcluster (cf. \autoref{fig:teaser}(b)), the tool fits the EBM model (precision and recall: $1.00$; see \autoref{fig:teaser}(a.3)) to this selection of points (against the rest) and recomputes feature importance locally. Insulin and SkinThick are the two most important dimensions and require further examination (see \autoref{fig:teaser}(c)). When selecting the right subcluster, the feature ranking produces the opposite result (SkinThick first and then Insulin). She acknowledges that these two dimensions are by far the most important for the formation of this entire cluster.

Sara clicks on the first feature, which is \emph{Insulin} (\autoref{fig:teaser}(c), green box), and the Score and Density graphs appear in~\autoref{fig:teaser}(d.1 and d.2). In \autoref{fig:teaser}(d.1), negative Score, she notices that the values of the selected points for Insulin are very low with most entries in the first bucket [$0$-$52.9$] (\autoref{fig:teaser}(d.2)). This feature follows a linear trend with high uncertainty (see the wide, gray band) for the rest of the data points, as shown in~\autoref{fig:teaser}(d.1), positive Score. However, the same effect is observable for the right subcluster (omitted due to space limits). She decides to continue with the exploration of the second important feature, which is SkinThick.
%
She checks the individual feature importance of SkinThick for the left and right subclusters, as presented in \autoref{fig:subclusters}, (a) and (b), respectively. She understands that the former contains mostly points with moderate values (cf. Density graph) and low uncertainty for this dimension (tight gray bar in Score graph; green box), while the latter demonstrates most points concentrated in values [$0$-$8.25$] with somewhat moderate uncertainty (see the green box in \autoref{fig:subclusters}(b)). Notably, Sara identified the opposite effect for SkinThick in the right subcluster compared to what she had seen in the inspection of the left subcluster. She concludes, then, that the common trait for this entire cluster is Insulin, while the reason behind its division into two subclusters is the different values in SkinThick. 


%% file: 5.Discussion.tex
\textbf{Design choices.} Although in its current version \textsc{DimVis} uses UMAP and EBM, the architecture of the tool allows it to be model-agnostic and flexible. The choices of EBM and UMAP were made since they are both state-of-the-art methods that can scale to large and high-dimensional datasets, but the flexible design of the workflow ensures its potential adaptability to emerging, future algorithms.
Also, one of the strengths of EBM is that it can be interpreted with simple visualizations of single or pairs of features (as opposed to, for example, visualizing decision trees as non-space-filling node-link diagrams). The visualizations used for \textsc{DimVis} are inherited from the EBM model (pre-evaluated, for example, with real-world medical data~\cite{caruana2015intelligible}) and do not represent a novelty of this paper. As such, a future direction could be to enhance the presentation beyond what is currently used. However, more complex visualizations may have a steeper learning curve and suffer in terms of scalability, effectiveness, and ease of use.

\textbf{Limitations and future work.} 
One important future work is an objective comparison to other similar tools and techniques so as to determine concretely how the detected explanations (e.g., features and their ranks) differ from other possible methods for generating similar explanations. This is one of our short-term plans for future work.
We also intend to refine the user experience based on feedback from domain experts (as in~\autoref{sec:case}) and perform a formal user study to validate our findings further and gather more feedback on possible improvements. Different possible visualization techniques (with variable degrees of complexity) will be tested with users to determine whether it is worth it to move beyond the current ones and how to do it.
We intend to investigate ways to improve the computational efficiency of the underlying model by, for example, using GPU-based implementations of both the DR and the EBM algorithms.
Finally, \textsc{DimVis} can currently only compare one data selection against the remaining projection, making a direct comparison between two subclusters a somewhat inefficient task. We intend to test the ability to select and compare two specific data subsets directly to streamline this process.

%% file: 6.Conclusion.tex
In this paper, we presented \textsc{DimVis}, a visualization tool using a supervised EBM model to interpret UMAP projections interactively. Our tool helps users with the in-depth exploration of cluster formations and their analysis based on the impact of single and pairwise features on the data. The applicability and usefulness of \textsc{DimVis} were tested with two use cases that showcase diverse settings for understanding real-world, high-dimensional data.